# Let a Thousand Flowers Bloom

## An Algebraic Representation for Edge Graphs

Jack Liell-Cock[a] and Tom Schrijvers[b]

a   University of Oxford, UK
b   KU Leuven, Belgium

**Abstract**
**Context**   Edge graphs are graphs whose edges are labelled with identifiers, and nodes can have multiple edges between them. They are used to model a wide range of systems, including networks with distances or degrees of connection and complex relational data.
**Inquiry**   Unfortunately, the homogeneity of this graph structure prevents an effective representation in (functional) programs. Either their interface is riddled with partial functions, or the representations are computationally inefficient to process.
**Approach**   We present a novel data type for edge graphs, based on total and recursive definitions, that prevents usage errors from partial APIs and promotes structurally recursive computations. We follow an algebraic approach and provide a set of primitive constructors and combinators, along with equational laws that identify semantically equivalent constructions.
**Knowledge**   This algebra translates directly into an implementation using algebraic data types, and its homomorphisms give rise to functions for manipulating and transforming these edge graphs.
**Grounding**   We exploit the fact that many common graph algorithms are such homomorphisms to implement them in our framework.
**Importance**   In giving a theoretical grounding for the edge graph data type, we can formalise properties such as soundness and completeness of the representation while also minimising usage errors and maximising re-usability.

## The Art, Science, and Engineering of Programming

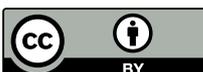

**Let a Thousand Flowers Bloom**

## 1 Introduction

Beyond illustrating thousands of flowers blooming, as demonstrated in Figure 1, graphs are a core programming tool that captures a range of computing structures and paradigms. Despite their universality, they continue to evade expression as a total algebraic data type. Common algebraic data types such as lists and trees have the liberty of a directional structure, as illustrated in Figure 2, which gives rise to a canonical inductive data type definition. Conversely, graphs benefit from homogeneity across their structure; no node or edge has priority over the others in the general case. This leaves many graph frameworks riddled with partial functions, verbose interfaces, or complex run times.

A common graph representation in computer systems is adjacency lists [7], but the internal consistency of this representation is not statically checked which leads to runtime errors. Erwig provides another alternative using inductive graphs and active patterns [4], but the interface also contains partial functions and some underlying operations are computationally inefficient – simply inserting a node takes $O(n \log n)$ time for dense graphs. Oliveira and Cook give a representation of cyclic graphs using fixpoints [12], but it has drawbacks of a non-intuitive portrayal, a lack of semantic equality testing, and difficulty in dynamically modifying the graphs. On the other hand, algebraic data types are expressed via total and recursive definitions. The lack of partiality prevents usage errors from invalid inputs and the recursive nature of the instantiation promotes structurally recursive computations. This also facilitates automated testing and proving properties by induction. It is therefore unsurprising that there have been additional attempts to represent graphs algebraically. Gibbons [5] provides a base instance for representing graphs as an algebraic interface, where they developed an initial algebra for constructing directed acyclic graphs, which captures certain recursive computation patterns. This work does not, however, accommodate undirected and cyclic graphs. More recently, Mokhov established an elegant algebra of node graphs [10]. It enjoys properties corresponding closely to those of a semiring and supports a rich algebraic graph library in Haskell that can be used to construct a wealth of complex graphs. Mokhov extended the research to support edge labels with a semiring structure [11]. A common theme among such extensions is this enforcement of structural constraints. Different semirings produce different graphs, but the additional structure restricts the nature of the edge labels and the versatility of the graph. This paper aims to overcome this drawback.

Instead of extending the node graph algebra to include edges, we invert the problem and develop an algebra for exclusively edge-indexed graphs, or edge graphs for short. In doing so, the variety of the edge labels is no longer restricted, and graphs that do not have node identifiers can be modelled. Our solution requires defining a novel graph representation that gives edge identifiers precedence over node identifiers. A set of constructors and equivalence relations encapsulating the structure is then defined, producing the edge graph algebra. The algebra contains six graph constructors, which are represented by the following algebraic data type:

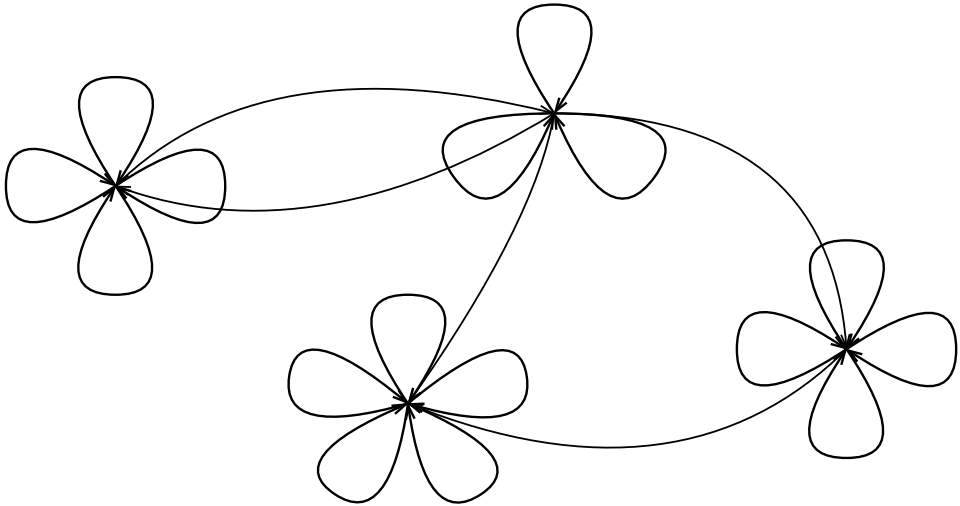

**■ Figure 1** Graph of flowers blooming

```
data EdgeGraph a =
  Empty
  | Edge a
  | Overlay (EdgeGraph a) (EdgeGraph a)
  | Into (EdgeGraph a) (EdgeGraph a)
  | Pits (EdgeGraph a) (EdgeGraph a)
  | Tips (EdgeGraph a) (EdgeGraph a)
```

Informally, *Empty* represents the empty graph while *Edge x* is a single edge with the label *x*. The *Overlay* constructor joins two graphs by unifying all nodes that have overlapping incoming or outgoing edges. *Into*, *Pits* and *Tips* each extend the *Overlay* operator in a different way. *Into* additionally connects each outgoing node in the first graph to each incoming node in the second. *Pits* connects each outgoing node in the first graph to each outgoing node in the second. And finally, *Tips* does the same as *Pits* but for incoming nodes. These operations are closed over the set of edge graphs and can be used to construct any edge graph. Moreover, each edge graph has a unique representation from the operations up to their equational laws which will be introduced later.

Given this unique representation of each graph, there is a unique mapping from an edge graph to any other collection of operators which also satisfies the edge graph axioms. Thus, graph algorithms may be translated to the problem of finding a model of the edge graph algebra that captures the algorithm's properties. We demonstrate this process with some key graph algorithms such as extracting the underlying set of edges, taking the transpose, and finding the shortest distances.

The rest of this paper is structured as follows. In Section 2 we outline previous work in the field of algebraic representations of graphs. In Section 3 we introduce typical graph representations along with our novel multigraph representation. The edge graph algebra is constructed in Section 4 and additional properties of the algebraic structure are shown. In Section 5 and 6, we provide Haskell implementations of the

**Let a Thousand Flowers Bloom**

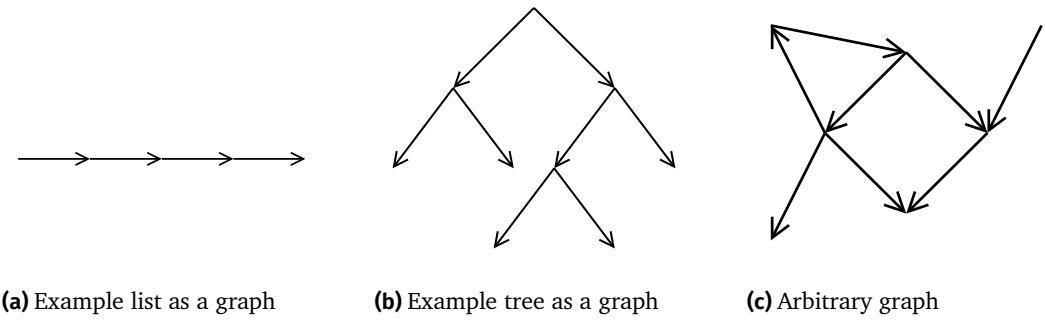

**(a)** Example list as a graph  **(b)** Example tree as a graph  **(c)** Arbitrary graph

■ **Figure 2** A range of common data types represented as graphs algebra, and express and implement graph algorithms in the form of homomorphisms. We outline future directions for this work and conclude in Section 7.

## 2 Literature Review

As outlined in Section 1, the literature is rich with attempts to characterise graphs with algebras. Many of the ideas in this paper build on previously presented ideas, thus a deeper recount of some key progressions in this field is warranted.

### 2.1 Directed Acyclic Multigraph Algebra

An algebra for directed acyclic multigraphs (which the paper refers to as a DAMG pronounced 'damage') was developed by Gibbons [5]. As it is a loaded phrase, we note to the reader that by the word *algebra*, we mean a collection of operators and constants (such as + and 0) and equational laws that all terms satisfy (such as $a + b = b + a$ and $a + 0 = a$). Concerning type definitions, the constructors are seen as the operators and constants, which in this case build graphs from smaller components. Rather than define their algebra via a representation, Gibbons takes a more abstract approach of letting the set of all graphs $D$ be the union of all $m, n \in \mathbb{N}$ indexed graphs $D_{m,n}$ which individually represent the set of DAMGs with $m$ inputs and $n$ outputs. With this definition, the constructors are given as:

1. The empty graph: $\varepsilon \in D_{0,0}$
2. A single edge: $e \in D_{1,1}$
3. A single vertex: $v_{m,n} \in D_{m,n}$
4. A *swap* graph: $s_{m,n} \in D_{m+n,n+m}$
5. A *beside* operator '|': for $x \in D_{m,n}$ and $y \in D_{p,q}$, then $x \mid y \in D_{m+p,n+q}$
6. A *before* operator '◁': for $x \in D_{m,n}$ and $y \in D_{n,p}$, then $x \triangleleft_n y \in D_{m,p}$

The interpretation of these constructors is that a vertex $v_{m,n}$ is a node with $m$ inputs and $n$ outputs, edges $e$ extend inputs or outputs, the *beside* operator composes graphs in parallel and the *before* operator composes graphs in series. We sometimes drop the subscript $n$ from the *before* operator when it is obvious from the context or is true for all $n$. Finally, the swap operator allows for graphs beyond the planar variety by connecting the first $m$ input edges with the last $m$ output edges and the last $n$ input edges with the first $n$ output edges.

Often with algebraic data types, different combinations of constructors can produce the same instantiation from a semantic viewpoint. Hence, one may introduce laws to quotient over equivalent constructions. There is usually little to enforce these equivalences in programming languages, so it is up to the programmer to ensure the functions treat semantically equal types the same. The set of axioms Gibbons introduces to identify equivalent graph constructions are:

- $(D, |, \varepsilon)$ is a monoid.
- $\triangleleft$ is associative (given it is correctly typed).
- The abiding law, $(a \triangleleft_m b) \mid (c \triangleleft_n d) = (a \mid c) \triangleleft_{m+n} (b \mid d)$.
- The swapping laws (where we say $m \times x = \underbrace{(x \mid \cdots \mid x)}_{m \text{ times}}$).
  - $s_{m,0} = m \times e$
  - $s_{m,n+p} = (s_{m,n} \mid (p \times e)) \triangleleft ((n \times e) \mid s_{m,p})$
  - $s_{m,n} \triangleleft (a \mid b) \triangleleft s_{p,q} = b \mid a$, where $a \in D_{n,p}, b \in D_{m,q}$

Gibbons showed that the algebra generated by these constructors and quotiented by the axioms is equivalent to an enriched symmetric strict monoidal category [8] and thus two DAMGs are semantically equal if and only if it can be proved so via the axioms [3]. Given this soundness and completeness of the axioms, the DAMG algebra is the free enriched strict symmetric monoidal category and thus there is a unique homomorphism to any other algebra that satisfies the DAMG algebra axioms. This unique mapping is called a catamorphism. In particular, a DAMG catamorphism $h$ is a tuple of constants, indexed constants and binary operators $(\!|w, x, y, z, \oplus, \otimes|\!)$ such that

$$h(\varepsilon) = w, \qquad h(v_{m,n}) = y_{m,n}, \qquad h(a \mid b) = h(a) \oplus h(b),$$
$$h(e) = x, \qquad h(s_{m,n}) = z_{m,n}, \qquad h(a \triangleleft b) = h(a) \otimes h(b),$$

where the constants and operators canonically form an enriched symmetric strict monoidal category. Immediate catamorphisms are the identity function $(\!|\varepsilon, e, v, s, |, \triangleleft|\!)$, a vertex count function $(\!|0, 0, 1, 0, +, +|\!)$, and a graph transpose function (reversing the direction of all of the edges) $(\!|\varepsilon, e, v, s', |, \triangleright|\!)$ where $s'_{m,n} = s_{n,m}$ and $a \triangleright b = b \triangleleft a$. A more interesting example given is a catamorphism which returns the length of the shortest path from entry to exit, which for the sake of brevity we point the reader to the original paper for the full specification.

## 2.2 Node Graph Algebra

The standard representation for node graphs is the relation representation $(N, E)$, where $N$ is the set of nodes and $E \subseteq N \times N$ is the set of edges. The interpretation of this representation is, an edge in the graph points from $n \in N$ to $m \in N$ if $(n, m) \in E$. We denote the collection of all such relation representations as $R$. Mokhov introduces an elegant algebra for graphs of this form [10].

The algebra is defined using four constructors, and their interpretation in the relational representation is as follows:

**Let a Thousand Flowers Bloom**

1. The empty graph: $\varepsilon = (\emptyset, \emptyset)$
2. A singleton node: $\dot{n} = (\{n\}, \emptyset)$
3. The *overlay* operator '+': $(N, E) + (N', E') = (N \cup N', E \cup E')$
4. The *connect* operator '$\gg$': $(N, E) \gg (N', E') = (N \cup N', E \cup E' \cup N \times N')$

The intuition behind the constructors is that the empty and singleton graphs provide the entry points into the graph construction. Then the *overlay* operator provides a way of discretely joining graphs while the *connect* operator provides a way of introducing connections between the nodes.

Mokhov introduces a minimal set of axioms that equate equivalent graphs with syntactically differing constructions.

- \+ is commutative and associative.
- $(R, \gg, \varepsilon)$ is a monoid.
- $\gg$ distributes over +.
- The decomposition axiom, $a \gg b \gg c = (a \gg b) + (a \gg c) + (b \gg c)$.

It can be deduced from these axioms that the + operator also has identity $\varepsilon$ and is idempotent. Thus, these axioms intriguingly reveal an algebraic structure similar to a semiring. The difference is the sharing of the identity by the two binary operators (preventing the annihilating *zero* element), and the additional decomposition axiom.

Beyond these core axioms, a range of other axioms can be introduced to recover common graph classes. In particular, the further axiom that $\gg$ is commutative gives the class of undirected graphs. Reflexive graphs can be attained with the additional axiom $\dot{n} = \dot{n} \gg \dot{n}$ specifically on nodes. Transitive graphs result from adding the axiom for all $b \neq \varepsilon$, $a \gg b + b \gg c = (a \gg b) + (a \gg c) + (b \gg c)$. Finally, the algebra can be extended to hypergraphs by replacing the decomposition axiom with an appropriate hyper-decomposition. For example, for 3-hypergraphs, the decomposition axiom becomes

$$a \gg b \gg c \gg d = (a \gg b \gg c) + (a \gg b \gg d) + (a \gg c \gg d) + (b \gg c \gg d).$$

## 2.3 Node and Semiring Edge Graph Algebra

The major setback of the node graph algebra developed by Mokhov [10] was the inability to identify edges, which is necessary for many graph algorithms. To overcome this drawback, Mokhov extended this research by introducing a Haskell tree type to model classes of graph algebras [11]. We present their work in a purely algebraic way to highlight the mathematical properties of their construction. Mokhov introduces an algebra $(N, \dot{\cdot}, \{\gg_e\}_{e \in E})$ where $N$ is the set of nodes, $\dot{\cdot}$ injects the node identifiers into singleton nodes in the graph, and $\gg_e$ is a set of edge-indexed binary operators where the set of edges has a semiring structure $(E, 0, 1, \oplus, \otimes)$.

The edge identifiers in Mokhov's algebra can come from an arbitrary semiring, which permits a form of edge labelling. The axioms imposed on the algebra are:

- $\gg_0$ is associative, commutative and idempotent
- Decomposition axioms, which implies associativity of $\gg_s$ for all $s \in e$
  - $(a \gg_s b) \gg_t c = (a \gg_s b) \gg_0 (a \gg_t c) \gg_0 (b \gg_t c)$,

- $a \gg_s (b \gg_t c) = (a \gg_s b) \gg_0 (a \gg_s c) \gg_0 (b \gg_t c)$.
- Parallel composition axiom, $(a \gg_s b) \gg_0 (a \gg_t b) = a \gg_{s \oplus t} b$.
- Transitivity axiom, $(a \gg_s b) \gg_0 (b \gg_t c) = (a \gg_s b) \gg_0 (a \gg_{s \otimes t} c) \gg_0 (b \gg_t c)$.
- Reflexivity axiom, $\dot{x} = \dot{x} \gg_1 \dot{x}$.

Many graph algorithms are equivalent to computing the transitive, reflexive closure of the graph edges over a given semiring. For example, Dijkstra's algorithm involves computing the closure of numerically labelled edges over the tropical semiring $(E, \infty, 0, \min, +)$ [9]. Hence, computing the transitive, reflexive closure of semiring edge-labelled graphs subsumes many graph algorithms. However, this requires forward knowledge of what is to be computed over the graph, and it does not allow for reinterpretation of the edge labels later. Moreover, the user may not want to process the graph via a semiring, and the algebra may induce confusion because the nodes are unique identifiers whereas the edges are potentially duplicated data labels. The algebra defined in our work does not have these drawbacks for the edge labels.

It is also worth noting that when $E$ is the boolean semiring $(\{0, 1\}, 0, 1, \wedge, \vee)$, the algebra aligns with the one for transitive, reflexive node graphs outlined in Section 2.2 without the empty graph, where $\gg_0 = +$ and $\gg_1 = \gg$. In particular, the containment law, $(a \gg_1 b) \gg_0 a \gg_0 b = a \gg_1 b$, is derivable from the parallel composition axiom, and distributivity of $\gg_1$ over $\gg_0$, is derivable from the decomposition axioms. The lack of the empty graph is made up by the containment law, which we will see in Section 2.4 is an equivalent formulation of the algebra that doesn't reference the empty graph. As one would expect, dropping the transitive and reflexive axioms recovers the original node graph algebra.

### 2.4 United Monoids

Mokhov also introduces an algebraic structure called the *united monoid* [11] which are a reoccuring concept in graph algebras. A united monoid $(X, \varepsilon, \oplus, \otimes)$ is a commutative monoid $(X, \varepsilon, \oplus)$ and a monoid $(X, \varepsilon, \otimes)$ such that $\otimes$ distributes over $\oplus$. That is, it is a semiring with identified units, without the annihilation axiom. Given this formulation, a commutative united monoid refers to the case where $\otimes$ is commutative. In any united monoid, $\oplus$ is idempotent,

$$a \oplus a = (a \otimes \varepsilon) \oplus (a \otimes \varepsilon) = a \otimes (\varepsilon \oplus \varepsilon) = a \otimes \varepsilon = a.$$

It also turns out that identifying the two units is equivalent to the containment axiom $(a \otimes b) \oplus a = a \otimes b$, which means that monoids can be united without reference to their identities, permitting the possibility of united semigroups. Some examples of united monoids in graph algebras and wider literature are given in [11].

## 3 Representations of Edge Graphs

In this section, we recall a range of common graph representations and discuss their inadequacy towards the problem we face. We necessarily introduce a new

**Let a Thousand Flowers Bloom**

graph representation for edge graphs and demonstrate its relation with the former representations. Throughout the rest of this paper, there is often a need to distinguish between the two ends of an edge. To avoid the polysemous term "tail", we say that an edge points from the *pit* to the *tip*. That is, edge $a$ points into edge $b$ if the tip of $a$ and the pit of $b$ share the same node. Similarly, an edge originates (or terminates) from a node if it shares its pit (or tip) with the node. Additionally, our edge graph representation will require numerous operations over pairs of sets, so we use the $-^2$ notation to "lift" a set operator to pairs of sets. For example, for $p_i$ the functions which project the $i$-th element of a tuple,

- $x \cup^2 y \stackrel{\text{def}}{=} (p_1(x) \cup p_1(y), p_2(x) \cup p_2(y))$,
- $x \subseteq^2 y \stackrel{\text{def}}{=} p_1(x) \subseteq p_1(y) \wedge p_2(x) \subseteq p_2(y)$,
- $\bigcup^2 X \stackrel{\text{def}}{=} \left(\bigcup\{p_1(x) \mid x \in X\}, \bigcup\{p_2(x) \mid x \in X\}\right)$.

### 3.1 Typical Graph Representations

The standard representation for node graphs is the relation representation $(N, E)$, outlined in Section 2.2. This representation can be extended to distinguish edges (and thus subsume multigraphs) by appending a set of edge identifiers $L$ to the edge pairs to form a triple $E \subseteq N \times N \times L$. This is an unsatisfying representation because it does not enforce that each edge identifier is used exactly once. So edges between two different pairs of nodes may have the same identifier, and the full set of edge identifiers may not be used in the graph, leaving superfluous data. A more complete representation of multigraphs is given by the tuple $(N, E, \pi, \tau)$, where $N$ is the set of nodes, $E$ the set of edges, and $\pi, \tau : E \to N$ select the pit and tip of each edge. We denote the set of all such multigraph representations $G$. This construction, however, still relies on the set of nodes to imbue the graph with relational information. Hence, drawbacks remain with this representation from the perspective of edge graphs. In particular, if the nodes are indistinguishable or vacant, this representation breaks down. Thus, the necessity of a representation independent of the nodes arises.

### 3.2 Flow Representation

In the relation representation, the set $N$ of nodes is explicit, and the edges are constructed secondarily from them. This definition is inverted to accommodate a graph with only edge identifiers. The edge graph is represented by an explicit set $E$ of edges which are used to construct the nodes. Each node is designated by a pair of subsets of $E$, corresponding to the node's associated pits and tips. The graph is then defined as a collection of these nodes adhering to some coherence conditions.

**Definition 1.** A flow representation for a set of edges $E$ is a subset $\gamma \subseteq \mathscr{P}(E) \times \mathscr{P}(E)$ such that
1. $\bigcup^2 \gamma = (E, E)$
2. $\forall x \neq y \in \gamma,\ x \cap^2 y = (\emptyset, \emptyset)$
3. $(\emptyset, \emptyset) \notin \gamma$

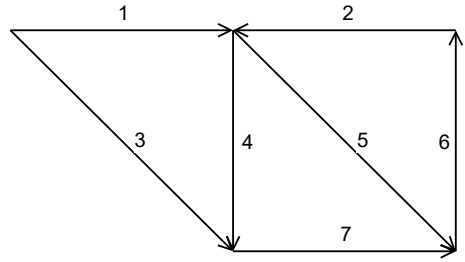

**Figure 3** A simple directed edge graph

where $\mathscr{P}$ is the powerset operator. The collection of all such flow representations is denoted $\Gamma$.

Informally, condition 1 says if an edge originates from a node, there better be a node that it terminates at (and visa versa). Moreover, the set of all of these originating and terminating edges is the full set $E$, so we don't have superfluous identifiers. Condition 2 states that an edge can only have its pit or tip coinciding with a single node: it doesn't make sense for an edge to be originating (or terminating) at more than one node. Finally, an isolated node – one without any tips or pits – is prohibited in an edge graph, so condition 3 stipulates that either the pits or tips of a node must be non-empty. The graph representation does not need to be augmented by the underlying set of edges because it is recoverable from the node representation. This representation is a coherent representation of edge graphs in that each edge graph has a unique flow representation and each flow representation gives rise to a unique edge graph. Further details are given in Theorem 6.

**Example 2.** The edge graph in Figure 3 has the flow representation

$$\gamma = \{(\emptyset, \{1,3\}), (\{1,2\}, \{4,5\}), (\{6\}, \{2\}), (\{3,4\}, \{7\}), (\{5,7\}, \{6\})\}.$$

For example, the bottom right node has incoming edges $\{5,7\}$ and outgoing edges $\{6\}$ which is captured by the last element $(\{5,7\},\{6\})$ of $\gamma$.

### 3.3 Nodal Flow Representation

The edge graph representation outlined in Section 3.2 can be canonically extended to include nodes. With this extension, an isomorphism with the multigraph representation presented in Section 3.1 is recovered, as stated in Theorem 4.

**Definition 3.** A nodal flow representation for a set of edges $E$ and nodes $N$ is the subset $\gamma \subseteq \mathscr{P}(E) \times \mathscr{P}(E) \times N$ such that
1. $\bigcup_{x \in \gamma} p_1(x) = E \ \wedge\ \bigcup_{x \in \gamma} p_2(x) = E \ \wedge\ \{p_3(x) \mid x \in \gamma\} = N$,
2. $\forall x \neq y \in \gamma,\ p_1(x) \cap p_1(y) = \emptyset \ \wedge\ p_2(x) \cap p_2(y) = \emptyset \ \wedge\ p_3(x) \neq p_3(y)$.

The collection of all such nodal flow representations is denoted $\Gamma^*$.

Note that the condition that either the set of tips or pits for a given node must be non-empty is no longer required because it is now possible to have isolated nodes.

**Let a Thousand Flowers Bloom**

Instead, we must specify that the full set of nodes is used in the graph, and no node identifier is used for two distinct nodes.

### 3.4 Coherence of the Flow Representations

To show the coherence of the flow representation for edge graphs, we first present the isomorphism between the nodal flow representation and the multigraph representation. We will then argue for an equivalence relation on the multigraph representation to quotient over graphs with analogous information from the perspective of an edge graph, and show that the flow representation is isomorphic to these equivalence classes of the multigraph representation. The proof of both isomorphisms is in the Appendix.

**Theorem 4.** *G is isomorphic to $\Gamma^*$.*

The benefit of the nodal flow representation is the ease of expression when the nodes are discarded because it does not rely on them for the relational information. The method of restricting a nodal flow representation to a flow representation is given by the function $f \circ (p_1 \times p_2)$, where $f(X) = \{x \in X \mid x \neq (\emptyset, \emptyset)\}$ and $p_i$ are the projections to the $i$-th elements of a tuple. Intuitively, each element of $\gamma$ is projected to remove the node, and any remaining pairs of empty pits and tips are filtered out.

**Definition 5.** Define equivalence $\sim$ by $(N, E, \pi, \tau) \sim (N', E, \pi', \tau')$ if there exists functions $\phi : \widetilde{N} \leftrightarrows \widetilde{N'} : \phi'$ such that $\phi \circ \pi|_{\widetilde{N}} = \pi'|_{\widetilde{N'}}$, $\phi \circ \tau|_{\widetilde{N}} = \tau'|_{\widetilde{N'}}$, $\psi \circ \pi'|_{\widetilde{N'}} = \pi|_{\widetilde{N}}$, and $\psi \circ \tau'|_{\widetilde{N'}} = \tau|_{\widetilde{N}}$, where $\widetilde{N} = \pi(E) \cup \tau(E) \subseteq N$, $\widetilde{N'} = \pi'(E) \cup \tau'(E) \subseteq N'$ and (without loss of generality) $\pi|_{\widetilde{N}}$ is the restriction of function $\pi : E \to N$ to $E \to \widetilde{N}$.

We argue that semantically equal multigraph representations from the perspective of edge graphs are captured by this equivalence relation. Informally, it identifies edge graphs up to the renaming of nodes that preserve the structure in the images of $\pi$, $\tau$, $\pi'$, and $\tau'$. The only nodes not in any of the images of these functions are isolated nodes and thus are irrelevant from the edge graph viewpoint. The following theorem validates the coherence of the flow representation.

**Theorem 6.** *$G/\sim$ is isomorphic to $\Gamma$.*

## 4 The Algebra

This section defines the algebraic structure of edge graphs. Algebraic characterisations are useful in the context of data types for proving properties and automated testing. The algebra consists of six constructors. Many combinations of these constructors correspond to the same graph which motivates the introduction of a set of equational laws to identify these semantically equivalent constructions. Similar to the graph algebras defined by Mokhov [11], the combinators form a set of united monoids. We additionally introduce a set of laws to describe the cohesion between these structures.

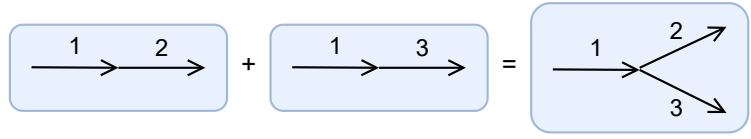

**Figure 4** Illustration of pairwise union

### 4.1 Constructors

Our method of building an abstract interface aims to give a canonical representation for every edge graph while maintaining simple axiom schemes that equate equivalent terms. Thus, the construction begins with the most basic graph, the empty graph $\varepsilon = \emptyset$. A graph with a single edge $e \in E$ is denoted by $\vec{e} \in \Gamma$ and has the flow representation $\{(\{e\},\emptyset),(\emptyset,\{e\})\}$. The two pairs represent the pit and tip of the single edge. These constructors are the base cases of the graph assembly.

Complexities arise in building the binary operators. In the node graph algebra, the binary operators were defined via unioning the components of the relation representation $R = (N, E)$. This is not feasible in the flow representation because it is not closed under unions. Hence, an operator is required to reduce the edge graphs by pairwise unioning all nodes that share an incoming or outgoing edge. We denote this operator with the symbol $+$. The following gives an example of an application of the operator which is also illustrated in Figure 4.

$$\{(\emptyset, \{1\}), (\{1\}, \{2\}), (\{2\}, \emptyset)\} + \{(\emptyset, \{1\}), (\{1\}, \{3\}), (\{3\}, \emptyset)\}$$
$$= \{(\emptyset, \{1\}), (\{1\}, \{2, 3\}), (\{2\}, \emptyset), (\{3\}, \emptyset)\}$$

This operator can be defined explicitly by realising that graph union on $R$ is the least upper bound on the partially ordered set (henceforth referred to as a poset) of node graphs with inclusion relation $\subseteq^2$. Hence, the objective is to enrich the edge graphs $\Gamma$ with an ordering relation too.

**Definition 7.** For two graphs $\gamma, \delta \in \Gamma$, $\gamma \preceq \delta$ if for all $x \in \gamma$, there exists a $y \in \delta$ such that $x \subseteq^2 y$.

This relation is reflexive, antisymmetric and transitive, so $(\Gamma, \preceq)$ defines a poset. Furthermore, for two edge graph representations $\gamma, \delta \in \Gamma$, let $\sim$ be a relation on the elements of $\gamma \cup \delta$ where $x \sim y \iff x \cap^2 y \neq (\emptyset, \emptyset)$. That is, two nodes are related if they share a pit or tip. This relation is reflexive and symmetric. Let $\sim^*$ be its transitive closure and define the join operator to be the set of unioned equivalence classes, $\gamma + \delta = \left\{ \bigcup^2 X \mid X \in \gamma \cup \delta / \sim^* \right\}$.

**Proposition 8.** $(\Gamma, \varepsilon, +)$ *is a bounded join-semilattice.*

*Proof.* We show that $\gamma + \delta$ is the least upper bound of $\gamma$ and $\delta$. Firstly, given $x \in \gamma$, let $[x] \in \gamma \cup \delta / \sim^*$ be its equivalence class. By definition, $x \subseteq^2 \bigcup^2[x]$ and $\bigcup^2[x] \in \gamma + \delta$. So $\gamma \preceq \gamma + \delta$. By the symmetry of the $+$ operator, $\delta \preceq \gamma + \delta$ too. Next, we must show that $\gamma + \delta$ is the least upper bound. Let $\zeta$ be another upper bound of $\gamma$ and $\delta$. Take some $X \in \gamma + \delta/\sim^*$, so for all $x \in X$, there exists a $y \in \zeta$ such that $x \subseteq^2 y$. But all of these $x \in X$ have transitively overlapping pits or tips. So for $\zeta$ to be well defined, it must be a single $\hat{y} \in \zeta$ with $x \subseteq^2 \hat{y}$ for all $x \in X$. So $\bigcup^2 X \subseteq^2 \hat{y}$, and $\gamma + \delta \preceq \zeta$ as required. Finally, it is trivial that $\varepsilon$ is the least element of $(\Gamma, \preceq)$. □

Thus, the + operator joins two graphs by taking the simplest graph structure coherent with both arguments. So we define the *overlay* operator precisely as this.

**Definition 9.** The *overlay* operator, $\gamma + \delta$, is given by the least upper bound of $\gamma$ and $\delta$ in $(\Gamma, \preceq)$.

Operators *into*, *pits* and *tips*, denoted by $\gg$, $\diamond$ and $\times$, respectively, are defined using this least upper bound construction as well. Firstly, we define some intermediate helper functions $c_i, c_p, c_t : E \times E \to \Gamma$ which construct a graph of two edges connected from tip to pit, at the pits, and at the tips, respectively. Note, only the *into* operator needs to handle the case where the edges coincide because the output of $c_p$ and $c_t$ reduces correctly to a single edge in the flow representation (by the uniqueness of elements in sets) when the inputs are equal.

$$c_i(d,e) = \begin{cases} \{(\emptyset, \{d\}), (\{d\}, \{e\}), (\{e\}, \emptyset)\} & \text{if } d \neq e \\ \{(\{d\}, \{d\})\} & \text{if } d = e \end{cases}$$

$$c_p(d,e) = \{(\emptyset, \{d,e\}), (\{d\}, \emptyset), (\{e\}, \emptyset)\}$$

$$c_t(d,e) = \{(\emptyset, \{d\}), (\emptyset, \{e\}), (\{d,e\}, \emptyset)\}$$

These functions, along with the underlying function $|-| : \Gamma \to E$ which takes the underlying edges of a graph, allow for the definition of the *into*, *pits* and *tips* operators.

**Definition 10.** The *into*, *pits* and *tips* operators are given by

$$\delta \gg \gamma \stackrel{\text{def}}{=} \delta + \gamma + \sum_{d \in |\delta|} \sum_{e \in |\gamma|} c_i(d,e),$$

$$\delta \diamond \gamma \stackrel{\text{def}}{=} \delta + \gamma + \sum_{d \in |\delta|} \sum_{e \in |\gamma|} c_p(d,e),$$

$$\delta \times \gamma \stackrel{\text{def}}{=} \gamma + \delta + \sum_{d \in |\delta|} \sum_{e \in |\gamma|} c_t(d,e).$$

For convenience, precedence order is given as *pits*, *tips*, *into* and finally *overlay*, e.g. $a + b \diamond d \gg d$ is equivalent to $a + ((b \diamond c) \gg d)$.

**Example 11.** Some simple edge graphs which are illustrated in Figure 5 are:
1. $\vec{1} + \vec{2}$ is the graph with two isolated edges
2. $\vec{1} \gg \vec{2}$ is the graph with edge 1 pointing into edge 2
3. $\vec{1} \diamond \vec{2}$ is the graph with two edges joined at the pit
4. $\vec{1} \times \vec{2}$ is the graph with two edges joined at the tip
5. $\vec{1} \gg \vec{1}$ is a petal graph – a single edge looped on itself
6. $(\vec{1} + \vec{2}) \gg (\vec{3} + \vec{4})$ is a cross graph, with two edges terminating at the node the other two edges originate from

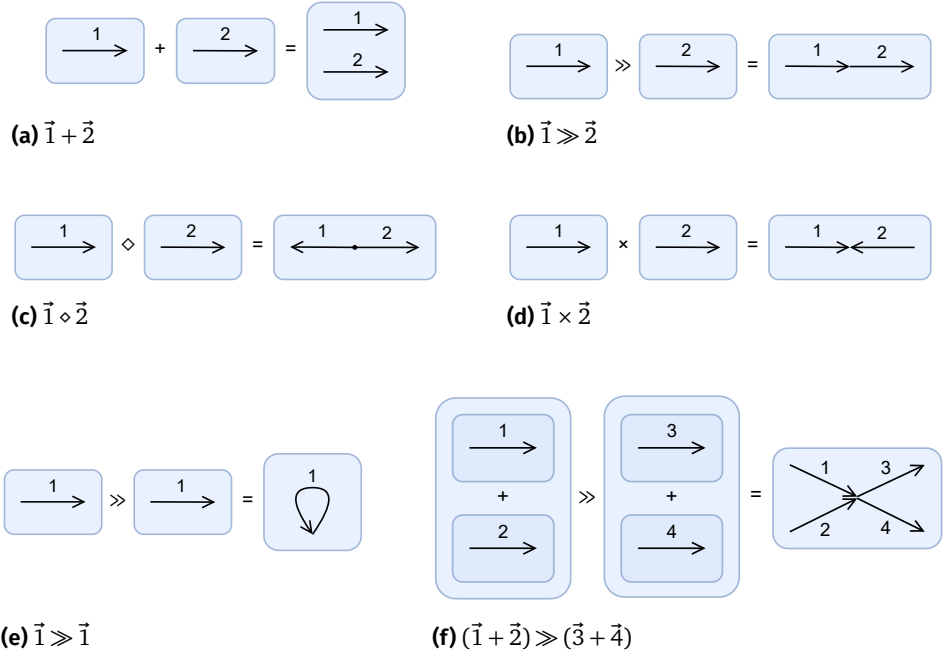

**Figure 5** Some simple edge graph constructions

## 4.2 The Axioms

As alluded to earlier, many constructions using the given primitives correspond to the same graph. For example, $\vec{1} = \vec{1} + \varepsilon$ and $\vec{1} + \vec{2} = \vec{2} + \vec{1}$. This section completes the definition of the edge graph algebra by developing a set of equational laws to identify such equivalences.

Verifying that $(\Gamma, +, \varepsilon)$, $(\Gamma, \gg, \varepsilon)$, $(\Gamma, \diamond, \varepsilon)$ and $(\Gamma, \times, \varepsilon)$ are all monoids is straightforward from the definitions. Moreover, so is that $+$, $\diamond$ and $\times$ are commutative and $+$ is idempotent. Hence, the elements of $(\Gamma, +, \varepsilon)$ are isomorphic to the sets of the underlying edges $E$. In fact, the latter three monoids are all united with the $(\Gamma, +, \varepsilon)$ monoid in the sense of Section 2.4. For this reason, we call the set of binary operators excluding the $+$ operator the *connect* operators. The association to the *connect* operator in Mokhov's original node graph algebra [10] is deeper than just the name which we plan to explore in a later paper. Furthermore, decomposition schemas similar to the ones defined for semiring edge graphs [11] hold:

$$a \square (b \blacksquare c) = a \square b + a \square c + b \blacksquare c$$
$$(a \square b) \blacksquare c = a \square b + a \blacksquare c + b \blacksquare c$$

where $\square$ and $\blacksquare$ are any of the *connect* operators. In the case where $\square$ and $\blacksquare$ represent the same operator, this schema generalises the associativity of $\gg$, $\diamond$ and $\times$.

These three united monoids and the decomposition schemas are almost sufficient to quotient the edge graph algebra to align with the intuitive description given in Section 4.1. All that is left is to define the relationships between the united monoids.

**Let a Thousand Flowers Bloom**

### 4.2.1 The Transitive Axioms
The first set of additional axioms comes from the understanding that three edges meeting at a single node may be constructed by connecting any two different pairs of the three edges. That is, $\vec{1} \diamond \vec{2} + \vec{2} \diamond \vec{3} = \vec{1} \diamond \vec{2} + \vec{1} \diamond \vec{3}$. These equivalent constructions can be identified by equating each pairwise construction to its transitive form. That is, for all $a \neq \varepsilon$,

$$a \diamond b + a \diamond c = a \diamond b \diamond c,$$
$$b \gg a + a \diamond c = b \gg a \diamond c,$$
$$a \gg b + a \gg c = a \gg b \diamond c,$$
$$a \times b + a \gg c = a \times b \gg c,$$
$$b \gg a + c \gg a = b \times c \gg a,$$
$$a \times b + a \times c = a \times b \times c.$$

The specification that $a$ is non-empty is because the transitive connection must be via one or more edges. Therefore, the empty graph, which has no edges, cannot act as this link. Otherwise, each axiom would reduce the $\diamond$ or $\times$ operator to be equivalent to the $+$ operator. For example, if $a = \varepsilon$ in the first axiom, this would result in

$$b + c = \varepsilon \diamond b + \varepsilon \diamond c = \varepsilon \diamond b \diamond c = b \diamond c.$$

### 4.2.2 The Reflexive Axioms
The astute reader will observe there is one final set of constructions using the graph operators which is equivalent but not provably so via the previous axioms. When a single edge is merged with itself via the $\diamond$ (or $\times$) operator, its one pit (or tip) is combined with itself and thus has no effect.

$$\vec{x} \diamond \vec{x} = \vec{x} + \vec{x} + c_p(x, x) = \vec{x} + \vec{x} + \vec{x} = \vec{x}$$

This equality can be instilled by a final pair of axioms, $\vec{x} \diamond \vec{x} = \vec{x}$ and $\vec{x} \times \vec{x} = \vec{x}$.

### 4.2.3 All Together Now
As a brief overview, the axioms for the edge graph algebra are:
- $(\Gamma, \varepsilon, +, \gg)$ is a united monoid.
- $(\Gamma, \varepsilon, +, \diamond)$ is a commutative united monoid.
- $(\Gamma, \varepsilon, +, \times)$ is a commutative united monoid.
- All nestings of $\gg$, $\diamond$ and $\times$ decompose via $+$ (which implies $+$ is idempotent [10]).
- $\gg$, $\diamond$ or $\times$ operators may be merged via the six transitive axioms.
- $\vec{x} \diamond \vec{x} = \vec{x}$ and $\vec{x} \times \vec{x} = \vec{x}$

A tuple $(\Gamma, \varepsilon, \vec{\cdot}, +, \gg, \diamond, \times)$ which satisfies these axioms is called an *edge graph algebra*. These axioms together are enough to completely quotient the edge graph algebra to align with the intuitive description given in Section 4.1. These axioms give a sound and complete representation of edge graphs in the sense that if there is an equational proof that two terms in the algebra are equal they represent the same graph, and if two terms represent the same graph then there is an equational proof of this using the axioms. The proof of the following theorem along with some other notable laws derivable from the above axioms are given in the Appendix.

**Theorem 12** (Soundness and Completeness). *The flow representation is isomorphic to the free edge graph algebra* $(\Gamma, \varepsilon, \vec{\cdot}, +, \gg, \diamond, \times)$.

For some intuition behind this theorem, we give the canonical representation of the edge graph in Figure 3 in our edge graph algebra. That is, the result from passing $\gamma$ from Example 2 through this isomorphism.

**Example 13.** The edge graph in Figure 3 can be represented canonically in the edge graph algebra as

$$\gamma = \varepsilon \gg \vec{1} \diamond \vec{3} + \vec{1} \times \vec{2} \gg \vec{4} \diamond \vec{5} + \vec{6} \gg \vec{2} + \vec{3} \times \vec{4} \gg \vec{7} + \vec{5} \times \vec{7} \gg \vec{6}$$

### 4.3 Partial Order

Given the idempotent + operation, a partial order can be defined on the set of edge graphs.

**Definition 14.** $a \subseteq b \iff a + b = b$

As it turns out, this coincides with the poset relation $\preceq$ from Section 4.1 because $a \preceq b \iff a + b = b$ follows from the definition of $+$ via the least upper bound in $(\Gamma, \preceq)$. The following partial-order theorems also immediately follow from this definition via the idempotent operation:

- Least element: $\varepsilon \subseteq a$
- Binary operation order: $(a \subseteq a \square b) \wedge (a \subseteq b \square a)$ where $\square$ is any binary operator.
- Monotony: $a \subseteq b \implies (a \square c \subseteq b \square c) \wedge (c \square a \subseteq c \square b)$ where $\square$ is any binary operator.

### 4.4 Subtraction

To define graph subtraction, we introduce a set of *layered* edge graphs $\Gamma \setminus \Gamma = \{(a, b) \mid a, b \in \Gamma, b \subseteq a\}$ with objects a graph and subgraph pair. We denote these pairs as $a \setminus b \in \Gamma \setminus \Gamma$.

**Definition 15.** The partial order on layered graphs is defined by

$$a \setminus b \sqsubseteq c \setminus d \iff a \subseteq c \wedge d \subseteq b.$$

Informally, the layered graphs can be considered as the second graph destructively interfering with the first. Thus, there is a functor $\cdot \setminus \varepsilon : \Gamma \to \Gamma \setminus \Gamma$ which *freely* lifts a graph to a layered graph.

**Definition 16.** Graph subtraction is defined by the adjunction $(-) \dashv (\cdot \setminus \varepsilon)$, where $(-) : \Gamma \setminus \Gamma \to \Gamma$ is the functor defined by $a \setminus b \mapsto a - b$. That is,

$$a \subseteq b - c \iff a \setminus \varepsilon \sqsubseteq b \setminus c. \tag{1}$$

Similar reasoning can be used to define subtraction on the node graphs [10]. We recover $a - \varepsilon = a$ and $a - a = \varepsilon$ as expected in both cases.

## 5 Instantiations

There is no point in defining abstract algebra if we cannot instantiate our ideas into usable code. This section introduces the abstract interface for the edge graph algebra via a Haskell type class and outlines some implementation schemes for the class.

### 5.1 Abstract Interface

While the edge graph can be encapsulated by the data type *EdgeGraph* introduced in Section 1, we present a shallow embedding to maximise the usability of the code. The edge graph constructors can be defined in terms of a type class (so long as the type instances adhere to the axioms given in Section 4.2.3).

```
class EdgeGraph g where
  type Edge g :: *
  empty :: g
  edge  :: Edge g → g
  (+)   :: g → g → g    -- overlay
  (≫)   :: g → g → g    -- into
  (⋄)   :: g → g → g    -- pits
  (×)   :: g → g → g    -- tips
```

The associated type *Edge g* corresponds to the set of edges. The remaining functions correspond to the graph construction operations defined in Section 4.1. This interface immediately permits a wealth of utility functions for constructing graphs. Functions to convert a list of edges to a discrete graph, a flower graph, a graph with all edges sharing a pit, and a graph with all edges sharing a tip (illustrated in Figure 6a, 6b, 6c, and 6d, respectively) are given by:

```
discreteGraph, flowerGraph, pitGraph, tipGraph :: EdgeGraph g ⇒ [Edge g] → g
discreteGraph = foldr ((+) ∘ edge) empty
flowerGraph [] = empty
flowerGraph xs = foldr ((≫) ∘ edge) (edge (head xs)) xs
pitGraph = foldr ((⋄) ∘ edge) empty
tipGraph = foldr ((×) ∘ edge) empty
```

A graph with a single intersecting node can be constructed from a list of edges terminating at the node and a list of edges originating at the node, shown in Figure 6e.

```
intoGraph :: EdgeGraph g ⇒ [Edge g] → [Edge g] → g
intoGraph ts ps = tipGraph ts ≫ pitGraph ps
```

Finally, an arbitrary graph can be defined given a list of nodes (defined by a pair of lists containing the originating and terminating edges at each node) with the function:

```
type Node g = ([Edge g], [Edge g])
```

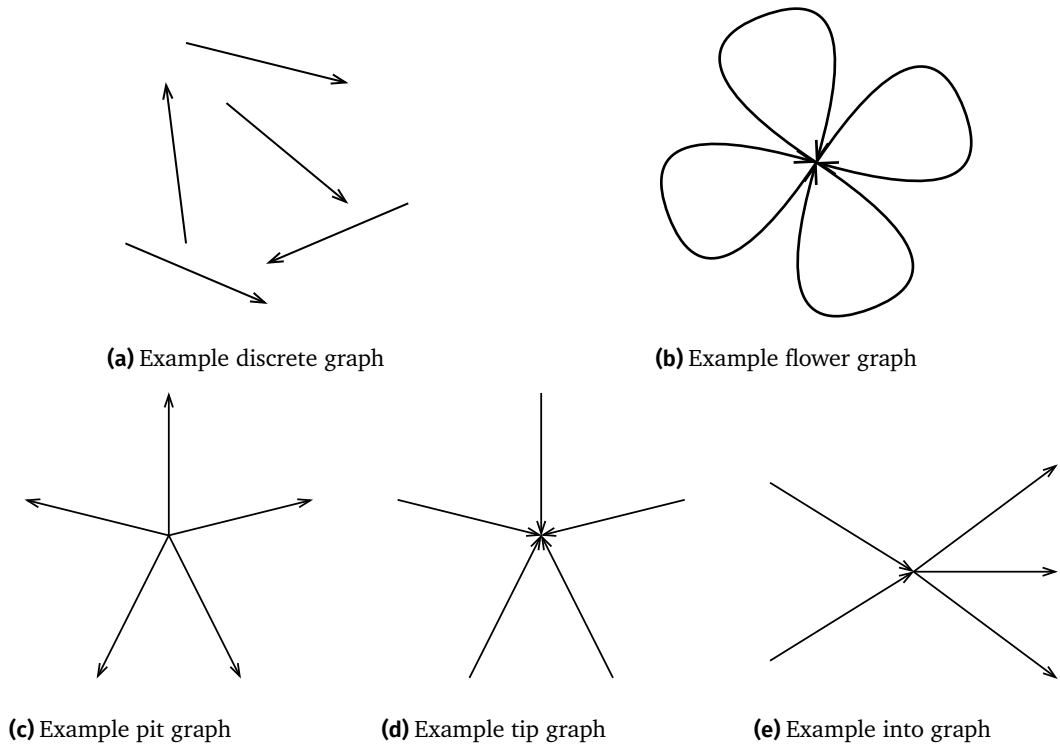

**(a)** Example discrete graph

**(b)** Example flower graph

**(c)** Example pit graph

**(d)** Example tip graph

**(e)** Example into graph

**Figure 6** A range of basic graph constructions

```
mkEdgeGraph :: EdgeGraph g ⇒ [Node g] → g
mkEdgeGraph = foldr ((+) ∘ uncurry intoGraph) empty
```

Even if the input list of pairs is ill-specified for a graph (i.e. two pairs might share the same pits or tips), the interface will automatically handle this by unifying the nodes as required. That is, all of the functions defined above are total and fully polymorphic which inhibits usage errors and maximises re-usability.

## 5.2 Deep Embedding

The edge graph algebra can be instantiated into a simple algebraic data type using a deep embedding.

```
data DeepGraph a =
    Empty
  | Edge a
  | DeepGraph a :+: DeepGraph a
  | DeepGraph a :≫: DeepGraph a
  | DeepGraph a :⋄: DeepGraph a
  | DeepGraph a :×: DeepGraph a
  deriving (Show)
instance EdgeGraph (DeepGraph a) where
  type Edge (DeepGraph a) = a
```

```
empty  = Empty
edge   = Edge
(+)    = (:+:)
(≫)    = (:≫:)
(⋄)    = (:⋄:)
(×)    = (:×:)
```

Semantics may be reinstated to the deep embedding by folding over it [2, 6].

```
foldShallow :: EdgeGraph g ⇒ DeepGraph (Edge g) → g
foldShallow Empty      = empty
foldShallow (Edge x)   = edge x
foldShallow (x :+: y)  = foldShallow x + foldShallow y
foldShallow (x :≫: y)  = foldShallow x ≫ foldShallow y
foldShallow (x :⋄: y)  = foldShallow x ⋄ foldShallow y
foldShallow (x :×: y)  = foldShallow x × foldShallow y
```

The data type *DeepGraph* does not yet fully capture the algebra because there is a syntactic difference between axiomatically equivalent graphs. Thus to implement *DeepGraph* as a function of the *Eq* type class, say, it must be reinterpreted to a representation where the syntactic structures are semantically equal – such as *FlowGraph* defined in Section 5.3:

```
instance Ord a ⇒ Eq (DeepGraph a) where
  g ≡ g' = foldShallow g ≡ (foldShallow g' :: FlowGraph a)
```

Much like the deep embedding of node graphs [10], the deep embedding of the edge graph can represent some graphs more compactly. The flow instantiation stores a duplicate copy of each edge to track the nodes associated with the pit and tip of the edge. The deep embedding is not restricted to this double trace and can, in the best case, store edge graphs with a single entry for each edge. However, the form of such edge graphs entails a low number of connection nodes. This often means most edges are discrete or form tight loops (i.e. a thousand flowers bloom). Thus, efficient deep representations are not very applicable given the impractical shape of edge graphs.

## 5.3 Flow Implementation

Another instantiation of the edge graph algebra follows the flow representation of edge graphs defined in Section 3.2. The underlying data type follows directly from the definition of $\Gamma$.[1]

```
data Node a = Node {tips :: Set a, pits :: Set a} deriving (Eq, Ord, Show)
type FlowGraph a = Set (Node a)
```

---

[1] In the following Haskell definitions, we import *Set* and *Map* modules fully qualified. We also note that the Haskell *Set* and *Map* libraries require an *Ord* constraint on the elements (or the keys in the case of *Map*) which is the primary reason for the *Ord* type constraints on the following functions.

Following the graph type above, we define additional utility functions that construct an empty node, take the union of two nodes, determine if two nodes share a common pit or tip, and embed a value into a node with a single pit or tip.

```
emptyN :: Ord a ⇒ Node a
emptyN = Node Set.empty Set.empty
unionN :: Ord a ⇒ Node a → Node a → Node a
unionN (Node ts ps) (Node ts' ps') = Node (Set.union ts ts') (Set.union ps ps')
disjointN :: Ord a ⇒ Node a → Node a → Bool
disjointN (Node ts ps) (Node ts' ps') = Set.disjoint ts ts' ∧ Set.disjoint ps ps'
pitN, tipN :: Ord a ⇒ a → Node a
pitN x = Node Set.empty (Set.singleton x)
tipN x = Node (Set.singleton x) Set.empty
```

The core of the implementation of *overlay* in the flow instantiation comes from a variation of the union-find algorithm, which aggregates all nodes that share the same pit or tip.

```
overlay :: Ord a ⇒ FlowGraph a → FlowGraph a → FlowGraph a
overlay = foldr f
   where f n acc = if null (tips n) ∧ null (pits n) then acc
      else Set.insert (foldr unionN n common) uncommon
         where (uncommon, common) = Set.partition (disjointN n) acc
```

Finally, the ≫, ⋄ and × implementations use the *overlay* algorithm along with their outline in Definition 10. First, the functions $c_i$, $c_p$, $c_t$, $|-|$ (or *und*) and a helper function *connect* are defined.

```
c_i, c_p, c_t :: Ord a ⇒ a → a → FlowGraph a
c_i x y
   | x ≡ y     = Set.singleton (Node (Set.singleton x) (Set.singleton x))
   | otherwise = Set.fromList [pitN x, Node (Set.singleton x) (Set.singleton y), tipN y]
c_p x y = Set.fromList [tipN x, tipN y, Node Set.empty (Set.fromList [x,y])]
c_t x y = Set.fromList [pitN x, pitN y, Node (Set.fromList [x,y]) Set.empty]
und :: Ord a ⇒ FlowGraph a → Set a
und = foldr (Set.union ∘ tips) Set.empty
connect :: Ord a ⇒ (a → a → FlowGraph a)
   → FlowGraph a → FlowGraph a → FlowGraph a
connect c g g' =
  let f x acc =
    let f' x' = overlay (c x x')
    in foldr f' acc (und g')
  in foldr f (overlay g g') (und g)
```

Then, the overall instantiation of the flow representation of the edge graph in Haskell is:

```
instance Ord a ⇒ EdgeGraph (FlowGraph a) where
  type Edge (FlowGraph a) = a
  empty = Set.empty
  edge x = Set.fromDistinctAscList [pitN x, tipN x]
  (+)  = overlay
  (≫)  = connect c_i
  (⋄)  = connect c_p
  (×)  = connect c_t
```

## 6 Catamorphisms

Theorem 12 showed the flow representation is isomorphic to the free edge graph algebra. That is, the flow representation forms the smallest algebra closed under the constructors in which all axioms from Section 4.2.3 hold. Therefore, there is a unique homomorphism to any other algebra that adheres to the axioms. We call this an edge graph catamorphism. In particular, a function $h : \Gamma \to B$ is an edge graph catamorphism if there exists a constant $e : B$, unitary function $f : E \to B$ and four binary functions $o, i, p, t : B \times B \to B$ such that

$$h(\varepsilon) = e, \qquad h(a+b) = o(h(a), h(b)), \qquad h(a \diamond b) = p(h(a), h(b)),$$
$$h(\vec{a}) = f(a), \qquad h(a \gg b) = i(h(a), h(b)), \qquad h(a \times b) = t(h(a), h(b)),$$

with these functions adhering to the axioms in Section 4.2.3 in the natural way. We write $(\!|e, f, o, i, p, t|\!)$ for such a set of functions defining a catamorphism.

**Examples of Catamorphisms** Some trivial examples of catamorphisms are immediate. The identity function on $\Gamma$ is $(\!|\varepsilon, \vec{\cdot}, +, \gg, \diamond, \times|\!)$. The underlying functor $|\!-\!|$, which reduces a graph to its underlying set of edges, is given by $(\!|\emptyset, \{\cdot\}, \cup, \cup, \cup, \cup|\!)$. The transpose of a graph is given by $(\!|\varepsilon, \vec{\cdot}, +, \ll, \times, \diamond|\!)$ where $a \ll b = b \gg a$.

An example beyond the trivialities is a shortest path function which returns the minimum distance to traverse between two edges if it is possible to navigate between them. For clarity, we express the fold via a Haskell implementation on the deep embedding. The algebra into which the edge graph will be folded over is the shortest paths algebra, *SP*, which is a map of the minimum distance between two pits/tips if they can be navigated via the edge graph. For efficiency, the fold is implemented over the product of the underlying and shortest paths algebra, because the underlying set is used in most of the shortest path algebra operators. In the end, the underlying set is discarded, hence this fold is a type of zygomorphism [13].

```
data End a = Pit a | Tip a deriving (Show, Eq, Ord)
type SP a = Map (End a, End a) a
type USP a = (Set a, SP a)
```

We define the *overlay* operator in this algebra by the *overlayUSP* function which simply unions the underlying edge set and unions over the shortest paths, taking the minimum path on conflicts. The *into*, *pits* and *tips* operators are defined via a generalised *connectUSP* function which assigns zero distance between a pit/tip in the first graph and a pit/tip in the second. The edge ends are configurable to allow the function to handle each *connect* operator case. These are not the complete algebra operators, however, because aggregating two graphs or identifying two nodes may produce new navigable paths. Hence *closureUSP* performs the transitive, reflexive closure on the *USP* type.

$$\begin{aligned}
&\textit{overlayUSP} :: \textit{Ord } a \Rightarrow \textit{USP } a \to \textit{USP } a \to \textit{USP } a \\
&\textit{overlayUSP } (s, m)\, (s', m') = (\textit{Set.union } s\, s', \textit{Map.unionWith min } m\, m') \\
&\textit{connectUSP} :: (\textit{Ord } a, \textit{Num } a) \Rightarrow \\
&\qquad (a \to \textit{End } a) \to (a \to \textit{End } a) \to \textit{USP } a \to \textit{USP } a \to \textit{USP } a \\
&\textit{connectUSP } e\ e'\ (s, m)\, (s', m') = \\
&\quad \mathbf{let}\ f\ x\ acc = \\
&\qquad \mathbf{let}\ g\ x'\ acc' = \\
&\qquad\quad \textit{Map.insertWith min } (e\,x, e'\,x')\ 0\ (\textit{Map.insertWith min } (e'\,x', e\,x)\ 0\ acc') \\
&\qquad \mathbf{in}\ \textit{foldr } g\ acc\ s' \\
&\quad \mathbf{in}\ (\textit{Set.union } s\ s', \textit{foldr } f\ (\textit{Map.unionWith min } m\ m')\ s) \\
&\textit{closureUSP} :: (\textit{Ord } a, \textit{Num } a) \Rightarrow \textit{USP } a \to \textit{USP } a \\
&\textit{closureUSP } (s, m) = \\
&\quad \mathbf{let}\ f\ (s, e)\ x\ acc = \\
&\qquad \mathbf{let}\ g\ (s', e')\ x'\ acc' \\
&\qquad\quad |\ e \equiv s' = \textit{Map.insertWith min } (s, e')\ (x + x')\ acc' \\
&\qquad\quad |\ \text{otherwise} = acc' \\
&\qquad \mathbf{in}\ \textit{Map.foldrWithKey } g\ acc\ m \\
&\qquad m' = \textit{Map.foldrWithKey } f\ m\ m \\
&\quad \mathbf{in\ if}\ m' \equiv m\ \mathbf{then}\ (s, m')\ \mathbf{else}\ \textit{closureUSP } (s, m')
\end{aligned}$$

Then *shortestPaths* algorithm can be defined succinctly as:

$$\begin{aligned}
&\textit{shortestPaths} :: (\textit{Ord } a, \textit{Num } a) \Rightarrow \textit{DeepGraph } a \to \textit{SP } a \\
&\textit{shortestPaths} = \textit{snd} \circ \textit{closureUSP} \circ h\ \mathbf{where} \\
&\quad h\ \textit{Empty}\ \ \ = (\textit{Set.empty}, \textit{Map.empty}) \\
&\quad h\ (\textit{Edge } x) = (\textit{Set.singleton } x, \\
&\qquad\qquad\qquad \textit{Map.fromList } [((\textit{Pit } x, \textit{Pit } x), 0), ((\textit{Pit } x, \textit{Tip } x), x), ((\textit{Tip } x, \textit{Tip } x), 0)]) \\
&\quad h\ (x :+: y)\ = \textit{overlayUSP } (h\,x)\,(h\,y) \\
&\quad h\ (x :\gg: y) = \textit{connectUSP Tip Pit } (h\,x)\,(h\,y) \\
&\quad h\ (x :\diamond: y)\ = \textit{connectUSP Pit Pit } (h\,x)\,(h\,y) \\
&\quad h\ (x :\times: y)\ = \textit{connectUSP Tip Tip } (h\,x)\,(h\,y)
\end{aligned}$$

Note that the current implementation uses the same value for the identifier and length of the edge. Hence, the algorithm cannot process a graph with two edges of equal length because they would be interpreted as the same edge. This could be solved by adding a function that maps the identifiers to a length to allow the identifiers to be unique, but for the sake of brevity and clarity, we omit this extension.

Additionally, the literal translation of the fold should apply *closureUSP* after each application of the *overlayUSP* or *connectUSP* functions. However, by realising that performing the closure of the shortest paths does not generate new information, applying *closureUSP* once at the end is equivalent to invoking it at every fold step. In effect, we are applying the fold fusion law in reverse. Hence this function is not an explicit fold, but it is more efficient and equivalent to one.

Finally, the shortest paths algorithm can be generalised to any semiring algorithm by letting *min* be the semiring sum, + the semiring product, and 0 the product identity. The unnavigable paths are implicitly given the semiring sum identity. Semiring algorithms solve a broad range of algebraic path problems – a large class of graph algorithms [1]. For example, the min-max path algorithm, which finds the path with the smallest maximal edge between two nodes, is also expressible via an edge graph algebra fold by replacing + with *max*; if edge weights represent fuel requirements between refills, this computes the fuel capacity required for a journey. Conversely, the max-min path algorithm solves the "low bridge" problem. We leave algorithm implementations beyond this class as future work but note that any edge graph algorithm is necessarily a homomorphism by the soundness and completeness of the algebra. Otherwise, the algorithm would produce different results for semantically equivalent graphs.

# 7 Conclusion

We have presented a novel representation for edge-indexed graphs that facilitates node-less presentations. We have formed a sound and complete algebra for this representation, and derived and implemented some natural graph algorithms as catamorphisms. An interesting question is whether this algebra may be generalised towards other graph algebras [10, 11] given the many shared themes between them. We are confident that our edge graph algebra can be extended to a hypergraph algebra by numerically labelling the ends of the graph edges instead of using the names *pits* and *tips*. We have the beginnings of an elegant formulation of the algebra in this format, but finding a canonical form and validating the axioms are sound and complete remains future work. We are hopeful this unified formal description of graphs can subsume this body of work.

**Acknowledgements**   We would like to thank Jeremy Gibbons for his invaluable discussions throughout the development of this work. The edge graph algebra would likely not exist without his early guidance and later reviews. Additionally, we are grateful to Sam Staton, Johannes Hartmann, Nicolas Wu and the anonymous reviewers for their feedback and helpful suggestions. Tom Schrijvers was partly funded by the FWO sabbatical bench fee K801223N.

## A  Proof of Graph Representation Isomorphisms

*Proof.* (Theorem 4). Firstly, given a multigraph representation $g = (N, E, \pi, \tau)$, we can construct a nodal flow representation by

$$\gamma = \{(\{e \in E \mid \tau(e) = n\}, \{e \in E \mid \pi(e) = n\}, n) \mid n \in N\}$$

This satisfies the first condition of the nodal flow representation. Taking the union over the first two projections amounts to the preimage of $\tau$ and $\pi$ for all of $N$ which is $E$. The union of the third projection is simply $N$ as required. The second restriction is similarly satisfied because it amounts to saying that the intersection of the preimages of $\tau$ and $\pi$ for distinct elements of $N$ are disjoint, which is true for any function.

Next, given a flow representation $\gamma$, we can construct $N$ and $E$ via

$$N = \{x_3 \mid (x_1, x_2, x_3) \in \gamma\},$$
$$E = \bigcup \{x_1 \mid (x_1, x_2, x_3) \in \gamma\} = \bigcup \{x_2 \mid (x_1, x_2, x_3) \in \gamma\}.$$

Then, because the set of first and second projections of the elements of $\gamma$ are disjoint covers of $E$, $\tau : E \to N$ can be constructed setting $\tau(e) = x_3$ for the unique $(x_1, x_2, x_3) \in \gamma$ such that $e \in x_1$. Similarly, $\pi : E \to N$ can be constructed by setting $\pi(e) = y_3$ for the unique $(y_1, y_2, y_3) \in \gamma$ such that $e \in y_2$. These two mappings between the multigraph and nodal flow representations are inverses of each other, hence $G \cong \Gamma^*$. □

*Proof.* (Theorem 6). Let $[(N, E, \pi, \tau)]$ be the equivalence class of the multigraph $g = (N, E, \pi, \tau)$. The goal is to construct a representative $\widehat{g} = (\widehat{N}, E, \widehat{\pi}, \widehat{\tau})$ of this equivalence class which has a one-to-one correspondence with an element from $\Gamma$. First, define $\widetilde{N} \subseteq N$ as above so it is the set of non-isolated nodes in $g$. As each non-isolated node corresponds to a unique pair of pits and tips, we can construct the isomorphism $\chi : \widetilde{N} \xrightarrow{\sim} \widehat{N}$ by

$$n \mapsto (\{e \in E \mid \tau(e) = n\}, \{e \in E \mid \pi(e) = n\}),$$

where $\widehat{N}$ is defined as the image of $\chi$. Let $\widehat{\pi} = \chi \circ \pi|_{\widetilde{N}}$ and $\widehat{\tau} = \chi \circ \tau|_{\widetilde{N}}$, so $\pi|_{\widetilde{N}} = \chi^{-1} \circ \widehat{\pi}$ and $\tau|_{\widetilde{N}} = \chi^{-1} \circ \widehat{\tau}$. As $\widehat{\pi} = \widehat{\pi}|_{\widehat{N}}$ and $\widehat{\tau} = \widehat{\tau}|_{\widehat{N}}$ by definition, letting $\phi = \chi$ and $\phi' = \chi^{-1}$ produces the equivalence $g \sim \widehat{g}$. A flow representation can be constructed from $\widehat{g} = (\widehat{N}, E, \widehat{\pi}, \widehat{\tau})$ simply by taking $\gamma = \widehat{N}$.

Next, from an arbitrary $\gamma \in \Gamma$, a multigraph equivalence class $[(N, E, \pi, \tau)]$ can be recovered by taking

$$N = \gamma,$$
$$E = \bigcup \{x_1 \mid (x_1, x_2) \in \gamma\} = \bigcup \{x_2 \mid (x_1, x_2) \in \gamma\}.$$

Then again set $\tau(e) = (x_1, x_2)$ for the unique $(x_1, x_2) \in \gamma$ such that $e \in x_1$ and $\pi(e) = (y_1, y_2)$ for the unique $(y_1, y_2) \in \gamma$ such that $e \in y_2$. This reconstructs a multigraph of the form of the representative $\widehat{g}$ in the equivalence class, so these mappings between $G/\sim$ and $\Gamma$ are inverses of each other, and hence $G/\sim \cong \Gamma$. □

## B  Further Edge Graph Algebra Laws

A range of further laws can be constructed from the set of edge graph algebra axioms. We outline a few key ones.

### B.1  Extended Transitive Law

The extended transitive law states that any two overlayed graphs sharing a non-empty subgraph on the same side of an *into* operator may be merged via that subgraph. This is a key lemma used to prove the normal form of the edge graphs in Theorem 12.

**Lemma 17.** For $a \neq \varepsilon$,

$$a \times b \gg c + a \times d \gg f = a \times b \times c \gg d \diamond f,$$
$$b \gg a \diamond c + d \gg a \diamond f = b \times c \gg a \diamond d \diamond f.$$

*Proof.* We prove the first equality, the second follows similar reasoning. For $a \neq \varepsilon$,

$$a \times b \gg c + a \times d \gg f$$
$$= \quad \{ \text{ transitive axiom } \}$$
$$a \times b + a \gg c + a \times d + a \gg f$$
$$= \quad \{ \text{ transitive axioms } \}$$
$$a \times b \times d + a \gg c \diamond f$$
$$= \quad \{ \text{ transitive axiom } \}$$
$$a \times b \times d \gg c \diamond f$$

$\square$

### B.2  Self-Loop Law

The self-loop law states that if a graph is looped onto itself, it can move across the $\gg$ operator.

**Proposition 18.** $a \diamond b \gg c + a \gg a = b \gg a \times c + a \gg a$

*Proof.* The proof for $a = \varepsilon$ is trivial. For non-empty $a$,

$$a \diamond b \gg c + a \gg a$$
$$= \quad \{ \text{ decomposition axiom } \}$$
$$a \diamond b + a \gg c + b \gg c + a \gg a$$
$$= \quad \{ \text{ transitive axiom } \}$$
$$a \diamond b + a \gg c \times a + b \gg c$$
$$= \quad \{ \text{ transitive axiom } \}$$
$$a \diamond b \gg c \times a + b \gg c$$
$$= \quad \{ \text{ transitive axiom } \}$$

$$a \diamond b \gg a + a \times c + b \gg c$$
$$= \quad \{ \text{ transitive axiom } \}$$
$$a \gg a + b \gg a + a \times c + b \gg c$$
$$= \quad \{ \text{ decomposition axiom } \}$$
$$b \gg a \times c + a \gg a$$

All of the transitive applications are over $a$, or a graph containing $a$, and thus are over non-empty graphs as required. □

### B.3 Containment Laws

The containment laws are properties of united monoids. The proof follows from the united units and distributivity [11].

**Proposition 19.** *For $\square$ any of $\gg$, $\diamond$ or $\times$,*

$$a \square b = a \square b + a = a \square b + b = a \square b + a + b.$$

## C  Proof of Soundness and Completeness

Here we give the proof of the soundness and completeness for the edge graph algebra axioms. We first prove a simple lemma regarding the edge graph ordering which will be used in the main proof. Lemma 17 from the previous section is also required.

**Lemma 20.** For an edge graph $\gamma \in \Gamma$, if $x \in |\gamma|$, then $\gamma + \vec{x} = \gamma$.

*Proof.* For a graph $\gamma \in \Gamma$, $x \in |\gamma|$ implies $\vec{x} \subseteq \gamma$, and the result follows immediately from the definition of the order. □

*Proof.* (Soundness and Completeness). The mapping of the flow representation to the algebra is simple. Given a flow graph representation $\gamma$, the algebraic representation can be formulated as

$$\sum_{(T,P)\in\gamma} \left( \mathcal{T}_{t\in T} \vec{t} \gg \prod_{p\in P} \vec{p} \right), \tag{2}$$

where $\sum$, $\mathcal{T}$ and $\prod$ are the reductions for $+$, $\diamond$ and $\times$, respectively. In the other direction, any algebraic graph representation can be converted into the above format (from which the flow representation can be easily extracted) via the following steps. We apply the conversion steps to an example graph,

$$\vec{1} \diamond (\vec{2} \gg \vec{3}) + \vec{1} \times (\vec{4} \diamond \vec{5} + \varepsilon) + \vec{3} \diamond \vec{1} + \vec{6} \times \vec{6},$$

along the way for clarity.

1. Collapse any $\varepsilon$ as it is unital to all binary operators. If only a single $\varepsilon$ remains, the conversion is complete giving $\gamma = \emptyset$ in the above representation.

$$\vec{1} \diamond (\vec{2} \gg \vec{3}) + \vec{1} \times (\vec{4} \diamond \vec{5}) + \vec{3} \diamond \vec{1} + \vec{6} \times \vec{6}$$

**Let a Thousand Flowers Bloom**

2. Use the distributivity and decomposition axioms to unnest the $\gg$, $\diamond$ and $\times$ operators to give edges joined by no more than one *connect* operator.
$$\vec{1} \diamond \vec{2} + \vec{1} \diamond \vec{3} + \vec{2} \gg \vec{3} + \vec{1} \times \vec{4} + \vec{1} \times \vec{5} + \vec{4} \diamond \vec{5} + \vec{3} \diamond \vec{1} + \vec{6} \times \vec{6}$$

3. Collapse duplicates using the idempotency of *overlay*, and *pits* and *tips* on singleton edges.
$$\vec{1} \diamond \vec{2} + \vec{1} \diamond \vec{3} + \vec{2} \gg \vec{3} + \vec{1} \times \vec{4} + \vec{1} \times \vec{5} + \vec{4} \diamond \vec{5} + \vec{6}$$

4. Apply $(\varepsilon \gg -)$ to any *pits* operations, $(- \gg \varepsilon)$ to any *tips* operations, and split singular edges, $\vec{x} = \varepsilon \gg \vec{x} + \vec{x} \gg \varepsilon$ using idempotency of $+$ and unital $\varepsilon$.
$$\varepsilon \gg \vec{1} \diamond \vec{2} + \varepsilon \gg \vec{1} \diamond \vec{3} + \vec{2} \gg \vec{3} + \vec{1} \times \vec{4} \gg \varepsilon + \vec{1} \times \vec{5} \gg \varepsilon + \varepsilon \gg \vec{4} \diamond \vec{5} + \varepsilon \gg \vec{6} + \vec{6} \gg \varepsilon$$

5. Use Lemma 17 to aggregate all *into* operators sharing an edge on the same side. Note that some of the variables in the lemma statement may be implicitly empty.
$$\vec{2} \gg \vec{1} \diamond \vec{2} \diamond \vec{3} + \vec{1} \times \vec{4} \times \vec{5} \gg \varepsilon + \varepsilon \gg \vec{4} \diamond \vec{5} + \varepsilon \gg \vec{6} + \vec{6} \gg \varepsilon$$

6. For any edges appearing only once, use Lemma 20 to extract another singular copy.
$$\vec{2} \gg \vec{1} \diamond \vec{2} \diamond \vec{3} + \vec{3} + \vec{1} \times \vec{4} \times \vec{5} \gg \varepsilon + \varepsilon \gg \vec{4} \diamond \vec{5} + \varepsilon \gg \vec{6} + \vec{6} \gg \varepsilon$$

7. Apply $(- \gg \varepsilon)$ or $(\varepsilon \gg -)$ to the singular edges to put them on the *opposite* side of the $\gg$ operator than their counterpart.
$$\vec{2} \gg \vec{1} \diamond \vec{2} \diamond \vec{3} + \vec{3} \gg \varepsilon + \vec{1} \times \vec{4} \times \vec{5} \gg \varepsilon + \varepsilon \gg \vec{4} \diamond \vec{5} + \varepsilon \gg \vec{6} + \vec{6} \gg \varepsilon$$

8. This brings the algebraic representation to the form of Equation 2 and by construction each edge appears exactly once on either side of an *into* operator. Hence, inverting the formulation gives a coherent edge graph representation.
$$\{(\{2\}, \{1,2,3\}), (\{3\}, \emptyset), (\{1,4,5\}, \emptyset), (\emptyset, \{4,5\}), (\emptyset, \{6\}), (\{6\}, \emptyset)\}$$

These two mappings are well-defined and inverses of each other, so the algebraic representation quotiented by the axioms is isomorphic to the flow representation. □

**Let a Thousand Flowers Bloom**

## About the authors

**Jack Liell-Cock** is a computer science PhD student at the University of Oxford. His interests are in algebraic representations of computer systems. In particular, algebraic effects and concurrency.
**email**  jack.liell-cock@cs.ox.ac.uk
**www**  https://www.cs.ox.ac.uk/people/jack.liell-cock/
https://orcid.org/0009-0005-7121-8095

**Tom Schrijvers** is a professor of computer science at KU Leuven. His interests are in functional programming and programming language theory.
**email**  tom.schrijvers@kuleuven.be
**www**  http://people.cs.kuleuven.be/~tom.schrijvers
https://orcid.org/0000-0001-8771-5559